# An Efficient Human Visual System Based Quality Metric for 3D Video

Amin Banitalebi-Dehkordi, Mahsa T. Pourazad, and Panos Nasiopoulos

**Abstract** Stereoscopic video technologies have been introduced to the consumer market in the past few years. A key factor in designing a 3D system is to understand how different visual cues and distortions affect the perceptual quality of stereoscopic video. The ultimate way to assess 3D video quality is through subjective tests. However, subjective evaluation is time consuming, expensive, and in some cases not possible. The other solution is developing objective quality metrics, which attempt to model the Human Visual System (HVS) in order to assess perceptual quality. Although several 2D quality metrics have been proposed for still images and videos, in the case of 3D efforts are only at the initial stages. In this paper, we propose a new full-reference quality metric for 3D content. Our method mimics HVS by fusing information of both the left and right views to construct the cyclopean view, as well as taking to account the sensitivity of HVS to contrast and the disparity of the views. In addition, a temporal pooling strategy is utilized to address the effect of temporal variations of the quality in the video. Performance evaluations showed that our 3D quality metric quantifies quality degradation caused by several representative types of distortions very accurately, with Pearson correlation coefficient of 90.8 %, a competitive performance compared to the state-of-the-art 3D quality metrics.

## 1 Introduction

With the introduction of 3D technology to the consumer market in recent years, one of the challenges industry has to face is assessing the quality of 3D content and evaluating the viewer's quality of experience (QoE). While several accurate quality metrics have been designed for 2D content, still there is room for improvement when it comes to 3D video quality assessment and reliability and accuracy of 3D quality metrics. Assessing the quality of 3D content is much more difficult than that of 2D content. In the case of 2D, there are well-known factors such as brightness, contrast, and sharpness that affect perceptual quality. In the case of 3D, depth perception changes the impact that the above factors have on the overall perceived 3D video quality. Although the effect of these factors on 2D video quality has been extensively studied, we have a limited understanding of how these factors affect 3D perceptual quality. In addition, other factors such as the scene's depth range, display size and the technology used in 3D display (i.e., active or passive glasses, glasses-free auto-stereoscopic displays, etc.) - known as 3D quality factors - solely affect 3D video quality, while they have no effect on 2D video quality [1]-[3]. The study presented by Seuntiens [4] identifies some additional 3D quality factors such as "presence" and "naturalness" that also affect 3D perception. These factors are of particular interest nowadays, since they have potential relevance in the design and evaluation of interactive media. Chen et al. [5] suggest including two other 3D quality factors, "depth quantity" and "visual comfort", in the overall quality of 3D content. It has been shown that 3D quality factors have closer correlation with the overall 3D video quality compared to 2D quality factors [6][7]. Considering that the existing 2D quality metrics only account for 2D quality factors and ignore the effect of 3D quality factors such as depth and the binocular properties of the human visual system (HVS), using these metrics to evaluate the quality of 3D videos yields low correlation with subjective results [8]-[11]. This has been verified by applying existing 2D quality metrics on the right and left views separately, and averaging the values over two views, then comparing the results with subjective evaluations [8]-[11].

   To account for the effect of 3D quality factors when evaluating the quality of 3D content, Ha and Kim [12] proposed a 3D quality metric that is based only on the temporal and spatial disparity variations. However, the proposed metric does not include the effect of 2D-associated quality factors such as contrast and sharpness. It is known that the overall perceived 3D quality is dependent on both 3D depth perception (3D factors) and the general picture quality (2D factors) [1]-[3]. In [12] the Mean Square Error (MSE) is used as a quality measure, which is known to have low performance in accurately representing the human visual system [13][14]. Moreover, this method utilizes disparity information instead of the actual depth map, which may result in inaccuracies in the case of occlusions. In addition, when using disparity instead of depth, the same amount of disparity may correspond to different perceived depths, depending on the viewing conditions. Boev et al. categorized the distortions of 3D

This work was partly supported by Natural Sciences and Engineering Research Council of Canada (NSERC) under Grant STPGP 447339-13 and the Institute for Computing Information and Cognitive Systems (ICICS) at UBC.
M. T. Pourazad is with ICICS at the University of British Columbia and TELUS Communications Inc., Canada (email: pourazad@icics.ubc.ca). The other authors are with the Electrical Engineering Department and ICICS at the University of British Columbia, Vancouver, BC, Canada (e-mails: {dehkordi, panos}@ece.ubc.ca).

content to monoscopic and stereoscopic types and proposed separate metrics for each type of distortions [15]. In this approach, while the monoscopic quality metric quantitatively measures the distortions caused by blur, noise and contrast-change, the stereoscopic metric exclusively measures the distortions caused by depth inaccuracies. The main drawback of this approach is that it is unable to accurately measure the overall 3D quality, as it does not attempt to fuse the 2D and 3D associated factors into one index.

Considering that the '3D quality of experience' refers to the overall palatability of a stereo pair, which is not limited to image impairments, some quality assessment studies propose to combine a measure of depth perception with the quality of the individual views. The study in [16] formulates the 3D quality as an efficient combination of the depth map quality and the quality of individual views. The depth map quality is evaluated based on the error between the squared disparities of the reference 3D pictures and that of the distorted one and the quality of each view is measured via structural similarity (SSIM) index [13]. Another group of researchers proposes to form a 3D quality index by combining the SSIM index of individual views and the Mean Absolute Difference (MAD) between the squared disparities of the reference 3D content and the distorted one [17]. In both methods presented in [16] and [17], the quality of the individual views is directly used to assess the overall 3D quality. However, when watching 3D, the brain fuses the two views to a single mental view known as cyclopean view. This suggests incorporating the quality of cyclopean view instead of that of individual views in order to design more accurate 3D quality metrics [18][19]. To this end, Shao et al. [20] considered the binocular visual characteristics of HVS to design a full-reference image quality metric. In this approach, a left-right consistency check is performed to classify each view to non-corresponding, binocular fusion, and binocular suppression regions. Quality of each region is evaluated separately using the local amplitude and phase features of the reference and distorted views and then combined into an overall score [20]. In another work, Chen et al. propose a full-reference 3D quality metric, which assesses the quality of the cyclopean view image instead of individual right and left view images. In their approach the cyclopean view is generated using a binocular rivalry model [18]. The energy of Gabor filter bank responses on the left and right images is utilized to model the stimulus strength and imitate the rivalrous selection of cyclopean image quality. This work was later extended in [21] by taking into account various 2D quality metrics for cyclopean view quality evaluation and generating disparity map correspondences. The results of this study show that by taking into account the binocular rivalry in the objective 3D content quality assessment process, the correlation between the objective and subjective quality scores increases, especially in the case of asymmetrically distorted content [21]. Similarly, in the study by Jin et al. the quality of cyclopean view was taken into account in the design of a full-reference 3D quality metric for mobile applications, which is called PHVS-3D [19]. In this study the information of the left and right channels are fused using the 3D-DCT transform to generate a cyclopean view. Then, a map of local block dis-similarities between the reference and distorted cyclopean views is estimated using the MSE of block structures. The weighted average of this map is used as the PHVS-3D quality index. Although the proposed schemes in [18] and [19] take into account the quality of cyclopean view, they ignore the depth effect of the scene. The quality of cyclopean view on its own does not fully represent what is being perceived from watching a stereo pair as it only reflects the impairment associated with the cyclopean image. The overall human judgment of the 3D quality changes depending on the scene's depth level [4]-[7]. The impairments of 3D content are more noticeable if they occur in areas where depth level of the scene changes. Thus, a measure of depth quality in addition to the quality of cyclopean view has to be taken to account in the design of a 3D quality metric. One example of such approach is the full-reference PHSD quality metric proposed by Jin et al. [9]. Similar to PHVS-3D, PHSD fuses the information of the left and right views to simulate the cyclopean view and measure its quality. Then to take into account the effect of depth, the MSE between the disparity maps of the reference and distorted 3D content as well as the local disparity variance are incorporated. PHSD has been specifically designed for video compression applications on mobile 3D devices. Although PHSD has been proposed for evaluating the quality of 3D video, it does not utilize a temporal pooling strategy to address the effect of temporal variations of the quality in the video. The temporal pooling schemes map a series of fidelity scores associated to different frames to a single quality score that represents an entire video sequence [22]. Most of the existing full-reference stereoscopic quality metrics are either designed for 3D images or do not utilize temporal pooling.

In addition to the full-reference stereoscopic quality evaluation methods, there are also several no-reference stereoscopic quality assessment methods proposed in the literature that demonstrate competitive performance compared to full-reference metrics [23][24]. Note that the no-reference quality metrics try to estimate the perceptual quality of distorted content without using the information of the reference 3D content. In one of the recent works, Chen et al. proposed a no-reference quality assessment method for natural stereopairs that employs the binocular rivalry to extract the 2D and 3D quality features and train a support vector machine to measure the quality of stereopairs [23]. In another work by Ryu and Sohn [24], the authors proposed a no-reference quality assessment framework for stereoscopic images by modeling the binocular quality perception in the context of blurriness and blockiness [24]. Even though the above-mentioned stereoscopic quality metrics algorithms are no-reference, they still demonstrate high performance when compared to some full-reference quality metrics for 3D images [23][24].

In this paper, we propose a full-reference 3D quality metric, which combines the quality of the cyclopean view and the quality of depth map. In order to assess the degradation caused by 3D factors in the cyclopean view, a local image patch fusion method (based on HVS sensitivity to contrast) is incorporated to extract the local stereoscopic structural similarities. To this end, the information of the left and right channels is fused using the 3D-DCT transform. Then, we extract the local quality values using the structural similarity (SSIM) index to calculate the similarity between the reference cyclopean frame (fused left & right) and the distorted one. Moreover, the effect of depth on 3D quality of experience is taken into account through the depth map quality component by considering the impact of binocular vision on the perceived quality at every depth level. To this end, the variance of the disparity map and the similarity between disparity maps are incorporated to take to account the depth information in the proposed quality prediction model. HVS-based 2D metrics have been used in our design instead of MSE or MAD, as the former ones are reported to represent the perception of the human visual system more accurately [14][25]. A temporal pooling strategy is used to address recency and the worst section quality effects. Recency effect refers to the high influence of the last few seconds of the video on the viewer's ultimate decision on video quality. Worst section quality effect denotes the severe effect that the video segment with the worst quality has on the judgment of the viewers. The proposed metric can be tailored to different applications, as it takes into account the display size (the distance of the viewer from the display) and the video resolution. The performance of our proposed method is verified through extensive subjective experiments using a large database of stereoscopic videos with various simulated 2D and 3D representative types of distortions. Moreover the performance of our proposed scheme is compared with that of state-of-the-art 3D and 2D quality metrics in terms of efficiency as well as complexity.

The main contributions of this paper are summarized as follows: 1) Design and formulate quality measures for the cyclopean view and depth map and propose a 3D quality metric as a combination of these two quality measures that effectively predicts the quality of 3D content at the presence of different types of distortions, 2) Take into account the temporal quality effects, display size, and video resolution in the design of the 3D quality metric, 3) Create a large database of stereoscopic videos containing several different representative types of distortions that may occur during the multiview video compression, transmission, and display process, and 4) Verify the performance of the proposed quality assessment method through large-scale subjective tests and provide a comprehensive comparison with the state-of-the-art quality metrics.

The rest of this paper is organized as follows: Section 2 is dedicated to the description of the proposed metric, Section 3 describes our experiment setup, results and discussions are presented in Section 4, while Section 5 concludes the paper.

## 2 Proposed 3D quality metric

As mentioned in the Introduction section, the binocular perception mechanism of the human visual system fuses the two view pictures into a single so called cyclopean view image. In addition, the perceived depth, affects the overall perceptual quality of picture. Our proposed Human-Visual-system-based 3D (HV3D) quality metric takes into account the quality of the cyclopean view and the quality of the depth information. Cyclopean view quality component evaluates the general quality of cyclopean image, while the depth map quality component measures the effect of depth and binocular perception on the overall 3D quality. The cyclopean view is generated from the two view pictures (for the reference and distorted video) through the cyclopean view generation process. To mimic the binocular fusion of HVS, best matching blocks within the two views are found through a search process. These matching blocks are then combined in the frequency domain to generate the cyclopean view block. During the block fusion, Contrast Sensitivity Function (CSF) of HVS is taken into account through a CSF masking process. Then, the local similarities between the resulted cyclopean images are integrated into the overall cyclopean view quality component (Section 2.1). The depth map quality component is constructed by taking into account the quality of depth maps as well as the impact of depth variances (Section 2.2). As distortions can be perceived differently when there are different levels of depth present in a scene, depth map variances are computed as part of the depth map quality component of our method. The size of the blocks and geometry of the system structure in the depth map quality component are selected by considering the fovea visual focus in the HVS. Fig. 1 illustrates the flowchart of the proposed framework and the following subsections elaborate on our proposed method.

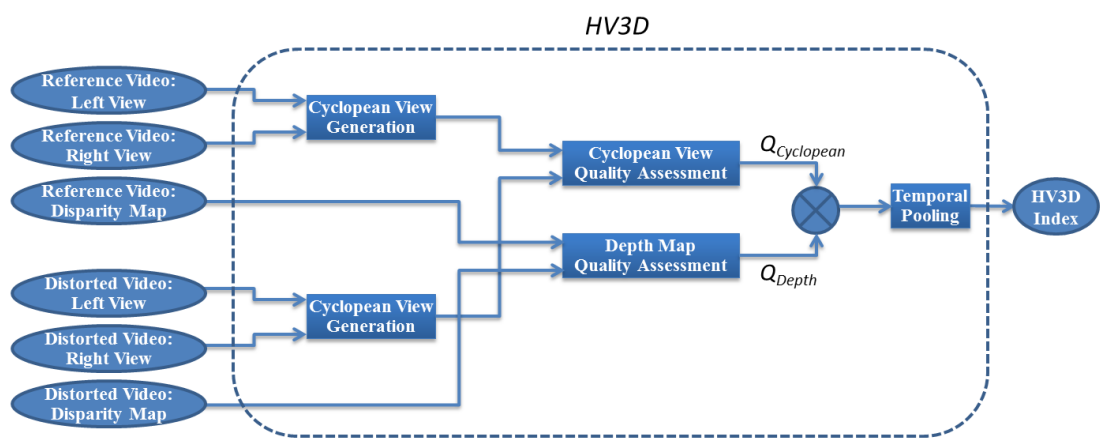

**Fig. 1** Illustration of the proposed quality assessment technique

*2.1 Quality of the cyclopean view – $Q_{Cyclopean}$*

To imitate the binocular vision and form the cyclopean view image we combine the corresponding areas from the left and right views. To this end, the luma information of each view is divided into $m \times m$ blocks. For each block of the left view, the most similar block in the right view is found, utilizing the disparity information. In our method, the depth map (or disparity map) is assumed to be available for the stereo views.

The depth map is either captured using a depth capturing device (e.g., a Kinect sensor), or generated by stereo matching techniques. There are many algorithms available in the literature for disparity map generation. We use the MPEG Depth Estimation Reference Software (DERS) [26] for dense disparity map generation. Note that we extract left-to-right disparity. The use of DERS for disparity map generation is suggested only when a dense disparity map is not available as an input. Note that in our cyclopean view generation scheme we need to find the best match to each block of pixels. However a dense disparity map provides the disparity value for each pixel. To address this problem, we approximate the disparity value of each block by taking the median of the disparity values of the pixels within that block. The depth information of each block has inverse relationship with its disparity value [27]. Having the disparity value per block, we can find the approximate coordinates of the corresponding block in the other view.

As Fig. 2 illustrates, $A_D$ is the approximate corresponding block for block $A_L$ in the right view, which has the same vertical coordinate as $A_L$ (illustrated as $j$ in Fig. 2), but its horizontal coordinate differs from $A_L$ by the amount of disparity (illustrated as $d$ in Fig. 2 ). Note that the $A_L$ and $A_D$ blocks are not necessarily the matching pair blocks that are fused by HVS since the position of $A_D$ is approximated based on the median of the disparity values of pixels within the block $A_L$. In the case of occlusions, the median value does not provide an accurate estimate of the block disparity, and can result to a mismatch.

To find the most accurate matching block, we apply a matching block technique based on exhaustive search in a defined $M \times M$ search range around $A_D$ (see Fig. 2) using the Mean Square Error cost function. In Fig. 2, $A_R$ is the best match for $A_L$ within the search range. Note that the block size and search range are chosen based on the display resolution. For instance, in the case of HD resolution video we choose *16×16* block size and the search area of *64×64*, since our performance evaluations have shown that these are the best possible sizes that significantly reduce the overall complexity of our approach while allowing us to efficiently extract local structural similarities between views.

The process of identifying matching blocks in the right view and the left view is done for both the reference and distorted stereo sets. Note that the distortions do not have any influence on the search results, since search is only performed on the reference video frames, and then the search results (i.e., the coordination of the best matching block) are used to identify matching blocks in the reference pair as well as the distorted pair.

In order to generate the cyclopean view, once the matching blocks are detected, the information of matching blocks in the left and right views needs to be fused. Here we apply the 3D-DCT transform to each pair of matching blocks (left and right views) to generate two $m \times m$ DCT-blocks, which contain the DCT coefficients of the fused blocks. The top level $m \times m$ DCT-block includes the coefficients of lower frequencies compared to the bottom one. Since the human visual system is more sensitive to the low frequencies of the cyclopean view [28], we only keep the $m \times m$ DCT-block corresponding to the lower frequency coefficients and discard the other ones. As a next step, we consider the sensitivity of the human visual system to contrast, which also affects the perceived image quality [29]. To take into account this HVS property, we need to prioritize the frequencies that are more important to the HVS. To this end, similar to the idea presented in [30], we utilize the proposed JPEG quantization tables. The JPEG quantization tables have been obtained from a series of psychovisual experiments designed to determine the visibility thresholds for the DCT basis functions. Based on this, the DCT coefficients that represent the frequencies with higher sensitivity to the human visual system are quantized less than the other coefficients so that the more visually important content is preserved during the course of compression. In our application, instead of quantizing the DCT coefficients we decided to scale them so that bigger weights are assigned to more visually important content. To achieve this, we adopt the 8×8 JPEG quantization table and create an 8×8 Contrast Sensitivity Function (CSF) modeling mask, such that the ratio among its coefficients is inversely proportional to the ratio of the corresponding elements in the JPEG quantization table. By applying the CSF modeling mask to the 3D-DCT blocks, the frequencies that are of more importance to the human visual system are assigned bigger weights. This is illustrated as follows:

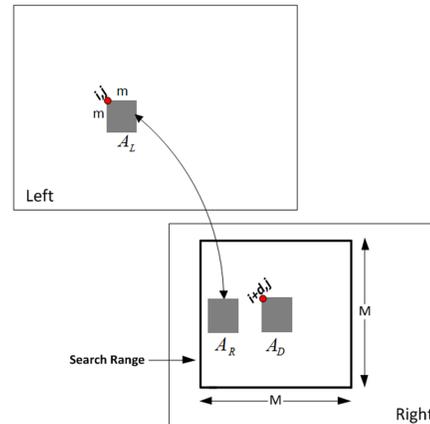

**Fig. 2** Selecting the best match: $A_D$ is the approximate corresponding block of $A_L$ in the right view + disparity; $A_R$ is the best match to $A_L$ within a search range

$$XC = C . \times X \quad (1)$$

where *XC* is our cyclopean-view model in the DCT domain for a pair of matching blocks in the right and left views, *X* represents the low-frequency 3D-DCT coefficients of the fused view, ".×" denotes the element-wise multiplication, and *C* is our CSF modeling mask. The elements of the CSF modeling mask are selected such that their average is equal to one. This guarantees that, in the case of uniform distortion distribution, the quality of each block within the distorted cyclopean view coincides with the average quality of the same view. Since the CSF modeling mask needs to be applied to $m \times m$ 3D-DCT blocks, for applications that require *m* to be greater than 8, cubic interpolation is used to up-sample the coefficients of the mask and create an $m \times m$ mask (in case *m* is less than 8, down-sampling will be used [31]).

Once we obtain the cyclopean-view model for all the blocks within the distorted and reference 3D views, the quality of the cyclopean view is calculated as follows:

$$Q_{Cyclopean} = \left( \sum_{i=1}^{N} \frac{SSIM(IDCT(XC_i), IDCT(XC'_i))}{N} \right)^{\beta_1} \quad (2)$$

where $XC_i$ is the cyclopean-view model for the $i^{th}$ matching block pair in the reference 3D view, $XC'_i$ is the cyclopean-view model for the $i^{th}$ matching block pair in the distorted 3D view, *IDCT* stands for inverse 2D discrete cosine transform, *N* is the total number of blocks in each view, $\beta_1$ is a constant exponent, and SSIM is the structural similarity index [13][14]. The value of $\beta_1$ is decided based on subjective tests presented in Section 3.

*2.2 Quality of the depth map – $Q_{Depth}$*

Depth information plays an important role in the perceptual quality of 3D content. The quality of the depth map becomes more important if there are several different depth levels in the scene. On the contrary, in a scene with a limited number of depth levels, the quality of the depth map plays a less important role in the overall 3D quality. This suggests considering the variance of depth map in conjunction with the depth map quality to reflect the importance of the depth map quality. However the variance of depth is required to be taken into account locally in the scene, as only a portion of the scene can be fully projected onto the eye fovea when watching a 3D display from a typical viewing distance. Fig. 3 illustrates the relationship between the block size projected onto eye fovea and the distance of the viewer. As it can be observed, the length of a square block on the screen that can be fully projected onto the eye fovea is calculated as follows:

$$K = 2 \times d \times \tan(\alpha) \quad (3)$$

where *K* is the length of the block (in [mm]), *d* is the proper viewing distance from the display (in [mm]), and *α* is the half of the angle of the viewer's eye at the highest visual acuity. The proper distance of a viewer from the display is decided based on the size of the display. The range of 2*α* is between 0.5° and 2° [32]. The sharpness of vision drops off quickly beyond this range. The length of the block (*K*) can be translated in pixel units as follows:

$$k = \frac{h \times K}{H} = \frac{2 \times d \times h \times \tan(\alpha)}{H} \quad (4)$$

where *k* is the length of the block on the screen (in pixels), *H* is the height of the display (in [*mm*]), and *h* is the vertical resolution of the display.

The local disparity variance is calculated over a block size area that can be fully projected onto the eye fovea when watching a 3D display from a typical viewing distance. For calculating the local depth-map variance of the $i^{th}$ block (i.e., $\sigma^2_{d_i}$), an outer block of $k\times k$ is considered such that the $m\times m$ block is located at its centre (see Fig. 4 for the present embodiment), and $\sigma^2_{d_i}$ is defined as follows:

$$\sigma^2_{d_i} = \frac{1}{k\times k - 1}\sum_{j,l=1}^{k}(M_d - R_{j,l})^2 \qquad (5)$$

where $M_d$ is the mean of the depth values of each $k\times k$ block (outer block around the $i^{th}$ $m\times m$ block) in the normalized reference depth map. The reference depth map has been normalized with respect to its maximum value in each frame, so that the depth values range from 0 to 1. $R_{j,l}$ is the depth value of pixel $(j,l)$ in the outer $k\times k$ block within the normalized reference depth map.

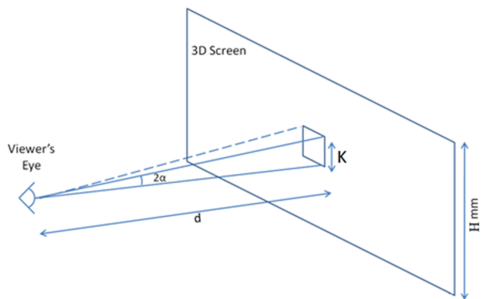

**Fig. 3** Visual acuity of fovea, receptive field; relationship between the block size projected onto eye fovea and the distance of the viewer

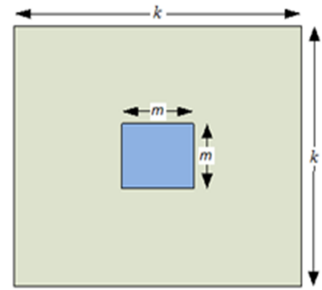

**Fig. 4** Block structures for calculating the variance of disparities

In our previous study on the relationship between the depth map quality and the overall 3D video quality, we verified that among the existing state-of-the-art 2D quality metrics, the Visual Information Fidelity (VIF) index of depth maps has the highest correlation with the mean opinion scores (MOS) of viewers [33]. It is also worth mentioning that the SSIM term in (2) compares the structure of two images only and does not take into account the effect of small geometric distortions [14]. The geometrical distortions are the source of vertical parallax, which causes severe discomfort for viewers (brain cannot properly fuse right and left view). The geometric distortions between right and left images (within the cyclopean view) are reflected in the depth map. Therefore, VIF is used to compare the quality of depth map of the distorted 3D content with respect to that of the reference one (see [14][25] for more details on VIF) as follows:

$$Q_{Depth} = (VIF(D,D'))^{\beta_2}\left(\sum_{i=1}^{N}\frac{\sigma^2_{d_i}}{N.\max(\sigma^2_{d_j}\,|\,j=1,2,...,N)}\right)^{\beta_3} \qquad (6)$$

where $D$ is the depth map of the reference 3D view, $D'$ is the depth map of the distorted 3D view, VIF is the Visual Information Fidelity index, $\beta_2$ and $\beta_3$ are constant exponents, $N$ is the total number of blocks, and $\sigma^2_{d_i}$ is the local variance of block $i$ in the depth map of the 3D reference view. Note that the latter part in equation (6) performs a summation over the normalized local variances. The variance term is computed for each block according to (5) and normalized to the maximum possible variance value for all the blocks. The summation is then divided by $N$, the total number of blocks, to provide an average normalized local variance value in the interval of [0,1].

*2.3 Constructing the overall HV3D metric*

Once the quality of the distorted cyclopean view and depth map is evaluated (see (2) and (6)), the final form of our HV3D quality metric is defined as follows:

$$HV3D = \left(\sum_{i=1}^{N}\frac{SSIM(IDCT(XC_i),IDCT(XC'_i))}{N}\right)^{\beta_1} \cdot (VIF(D,D'))^{\beta_2} \cdot \left(\sum_{i=1}^{N}\frac{\sigma^2_{d_i}}{N.\max(\sigma^2_{d_j}\,|\,j=1,2,...,N)}\right)^{\beta_3} \qquad (7)$$

where $XC_i$ is the cyclopean-view model for the $i^{th}$ matching block pair in the reference 3D view, $XC'_i$ is the cyclopean-view model for the $i^{th}$ matching block pair in the distorted 3D view, IDCT stands for inverse 2D discrete cosine transform, $N$ is the total number of blocks in each view, $\beta_1, \beta_2, \beta_3$ are the constant exponents, SSIM is the structural similarity index [13][14], $D$ is the depth map of the reference 3D view, $D'$ is the depth map of the distorted 3D view, VIF is the Visual Information Fidelity index, and $\sigma^2_i$ is the local variance of block $i$ in the depth map of the 3D reference view. The exponent parameters in HV3D ($\beta_1$, $\beta_2$, and $\beta_3$) are determined in the subsection 2.4.

Since different frames of a video have different influence on the human judgment of quality, the overall quality of a video sequence is found by assigning weights to frame quality scores, according to their influence on the overall quality. Subjective tests for video quality assessment have shown that subjects' ultimate decisions on the video quality are highly influenced by the last few seconds of the video (recency effect) [22][34][35]. Moreover, the video segment with the worst quality highly affects the judgment of the viewers [35]. This is true mainly because subjects keep the most distorted segment of a video in memory much more than segments with good or fair quality [35]. Temporal pooling algorithms have been proposed that map a series of fidelity scores associated to different frames to a single quality score that represents an entire video sequence [22]. To address the recency and worst section quality effects, a temporal pooling strategy has been used in our study, which is discussed in subsection 2.5. The overall approach proposed in our study to evaluate the quality of 3D content is illustrated as a flowchart in Fig. 1. A MATLAB implementation of our metric is available online at our web site [36].

*2.4 Constant exponents: $\beta_1$, $\beta_2$, and $\beta_3$*

To find the constant exponents for our HV3D quality metric and validate its performance, we performed subjective tests using two different 3D databases (one set for training and one set for validation).

To estimate the exponent constants of the proposed metric as denoted in the equation (7), all the terms were calculated for each video in the training dataset. To determine the best values for the exponent constants $\beta_1$ $\beta_2$, and $\beta_3$, we need to maximize the correlation between our HV3D indices and the MOS values of the training dataset. This can be formulated as follows:

$$\max_{\beta_i, i=1,2,3} \{\rho(HV3D, MOS)\} \tag{8}$$

where ρ is the Pearson correlation coefficient. We evaluate the correlation between HV3D and MOS vectors over a wide range of $\beta_i$ values and select the $\beta_i$ values that result in the highest correlation. The accuracy and robustness of the obtained exponents is further confirmed by measuring the correlation between MOS values and the HV3D indices of the validation video set (see Section 4).

*2.5 Temporal pooling strategy*

In our study, to address the recency and worst section quality effects we used the exponentially weighted Minkowski summation temporal pooling mechanism [37]. As shown in [22], the exponentially weighted Minkowski summation strategy outperforms some other existing temporal pooling methods, such as the histogram-based pooling, averaging, mean value of the last frames, and local minimum value of the scores in successive frames. The exponentially weighted Minkowski summation is formulated as follows:

$$HV3D_{\exp Minkowski} = [\frac{1}{N_f} \sum_{i=1}^{N_f} HV3D_i^p \cdot e^{\frac{i-N_f}{\tau}}]^{1/p} \tag{9}$$

where $N_f$ is the total number of frames, $p$ is the Minkowski exponent, and $\tau$ is the exponential time constant that controls the strength of the recency effect. Higher values of $p$ result in the overall score to be more influenced by the frame with largest degradation. In order to find the best values of $p$ and $\tau$, the Pearson correlation coefficient (PCC) is calculated for a wide range of these two parameters for the training video set.

Moreover, in order to adjust the proposed metric for the asymmetric video content where the overall quality of right and left view and their corresponding depth maps is not identical, the reference depth map and the base view for finding matching blocks (in the process of cyclopean view and depth map quality evaluation) are switched between the two views for every other frame. As a result, the overall 3D quality is not biased by the quality of one view or the eye dominance effect [38].

**3 Experiment setup**

This section provides details on the experiment setup, video sets used in our experiments, and the parameters of the proposed metric.

*3.1 Data set*

To adjust the parameters of our proposed scheme and verify its performance, we used two different video data sets (one training dataset and one validation dataset). The specifications of the training and validation video sets are summarized in Table 1 and Table 2, respectively. These sequences are selected from the test videos in [11], the 3D video database of the Digital Multimedia Lab (DML) at the University of British Columbia (publicly available [36]), and sequences provided by MPEG for standardization activities and subjective studies [39]. These datasets contain videos with fast motion, slow motion, dark and bright scenes, human and non-human subjects, and a wide range of depth effects. Note that the test sequences adopted from [11] and [36] are naturally captured stereoscopic videos, i.e., they contain only two views captured using two side-by-side cameras. The MPEG sequences, however, are originally multiview sequences with several views per each video. For each of the multiview sequences, only the two views (i.e., one stereoscopic pair) which were recommended by MPEG in their Common Test Conditions are used [39].

For each video sequence, the amount of spatial and temporal perceptual information is measured according to the ITU Recommendation P.910 [40] and results are reported in Table 1 and Table 2. For the spatial perceptual information (SI), first the edges of each video frame (luminance plane) are detected using the Sobel filter [41]. Then, the standard deviation over pixels in each Sobel-filtered frame is computed and the maximum value over all the frames is chosen to represent the spatial information content of the scene. The temporal perceptual information (TI) is based upon the motion difference between consecutive frames. To measure the TI, first the difference between the pixel values (of the luminance plane) at the same coordinates in consecutive frames is calculated. Then, the standard deviation over pixels in each frame is computed and the maximum value over all the frames is set as the measure of TI. More motion in adjacent frames will result in higher values of TI. Fig. 5 shows the spatial and temporal information indexes of each test sequence, as recommended in [40]. In addition to spatial and temporal information, for each sequence shown in Table 1 and Table 2, we also provide information about the scene's depth bracket. The depth bracket of each scene is defined as the amount of 3D space used in a shot or a sequence (i.e., a rough estimate of the difference between the distance of the closest and the farthest visually important objects from the camera in each scene) [42]. Since the information regarding the objects/camera coordinates is not available for all of the sequences, we adopt the disparity to depth conversion method introduced in [43] (which takes into account the display size and distance of the screen from the viewer) to find the depth of each object with respect to the viewer. We report the approximate averaged-over-frames of depth difference between the closest and farthest visually important objects. The visually important objects are chosen based on our 3D visual attention model [44] which takes into account various saliency attributes such as brightness intensity contrast, color, depth, motion, and texture. It is observed from Table 1 and Table 2 that the training and test videos have a similar distribution of properties (spatial and temporal complexity and depth bracket). This makes it possible to compute the required parameters in our proposed quality metric using the training video set and use the same ones for performance evaluations over the test video set.

In order to evaluate the performance of our proposed quality metric, in addition to the general distortions used by 2D quality metric studies [13][14][25] we take into account the distortions that may occur during 3D video content delivery and display. The suggested scheme for delivering 3D content is to transmit two or three simultaneous views of the scene and their corresponding depth maps, and synthesize extra views at the receiver end to support multiview screens [45]. In this process the delivered content might be distorted due to compression of views, compression of depth maps, or view synthesizing.

Once these distortions are applied to the content, the quality of videos is evaluated both subjectively and objectively using the HV3D metric and existing state-of-the-art 2D and 3D metrics. Note that the levels of distortions applied to the 3D content are such that they lead to visible artifacts which in turn allow us to correlate subjective tests - Mean Opinion Score (MOS) - with objective results.

The following types of distortions are applied to both training and validation 3D videos (seven different types of distortions applied to 16 original videos, which results in 208 distorted videos):

*1)* White Gaussian noise: white Gaussian noise with zero mean and variance value 0.01 is applied to both right and left views. DERS is used to generate a new depth map for the distorted stereo pair [26].

*2)* Gaussian low pass filter: a Gaussian low pass filter with the size 4 and the standard deviation of 4 is applied to both right

**Table 1** Training datasets

| Sequence | Resolution | Frame Rate (fps) | Number of Frames | Spatial Complexity (Spatial Information) | Temporal Complexity (Temporal Information) | Depth Range (cm) |
|---|---|---|---|---|---|---|
| Poznan_Hall2 | 1920×1080 | 25 | 200 | Low (35.4658) | Low (11.1460) | High (28.93) |
| Undo_Dancer | 1920×1080 | 25 | 250 | High (81.0423) | High (26.9021) | High (30.69) |
| Kendo | 1920×1080 | 30 | 300 | Medium (47.2172) | High (26.8791) | High (21.39) |
| Balloons | 1920×1080 | 30 | 500 | Medium (48.6726) | High (21.4660) | Low (5.84) |
| Cokeground | 1920×1080 | 30 | 210 | High (86.9096) | Medium (15.9128) | Low (4.99) |
| Ball | 1920×1080 | 30 | 150 | Medium (49.7701) | Low (13.3074) | Medium (15.53) |
| Alt-Moabit | 1920×1080 | 30 | 100 | High (111.0437) | High (21.2721) | Medium (13.36) |
| Hands | 1920×1080 | 30 | 251 | High (114.6755) | High (25.2551) | Medium (15.86) |

**Table 2** Validation datasets

| Sequence | Resolution | Frame Rate (fps) | Number of Frames | Spatial Complexity (Spatial Information) | Temporal Complexity (Temporal Complexity) | Depth Range (cm) |
|---|---|---|---|---|---|---|
| Poznan_Street | 1920×1080 | 25 | 250 | High (95.3103) | High (26.5562) | High (34.01) |
| GT_Fly | 1920×1080 | 25 | 250 | Medium (58.8022) | High (33.0102) | High (31.02) |
| Lovebird1 | 1920×1080 | 30 | 300 | Medium (59.2345) | Low (13.8018) | Medium (15.01) |
| Newspaper | 1920×1080 | 30 | 300 | High (65.1173) | Medium (17.1297) | Low (5.09) |
| Soccer2 | 1920×1080 | 30 | 450 | High (115.2781) | High (28.6643) | Medium (16.99) |
| Flower | 1920×1080 | 30 | 112 | Medium (43.0002) | Low (13.5305) | Low (5.86) |
| Horse | 1920×1080 | 30 | 140 | High (85.4988) | High (22.3184) | Medium (13.56) |
| Car | 1920×1080 | 30 | 235 | Medium (49.6162) | Medium (16.0197) | High (24.21) |

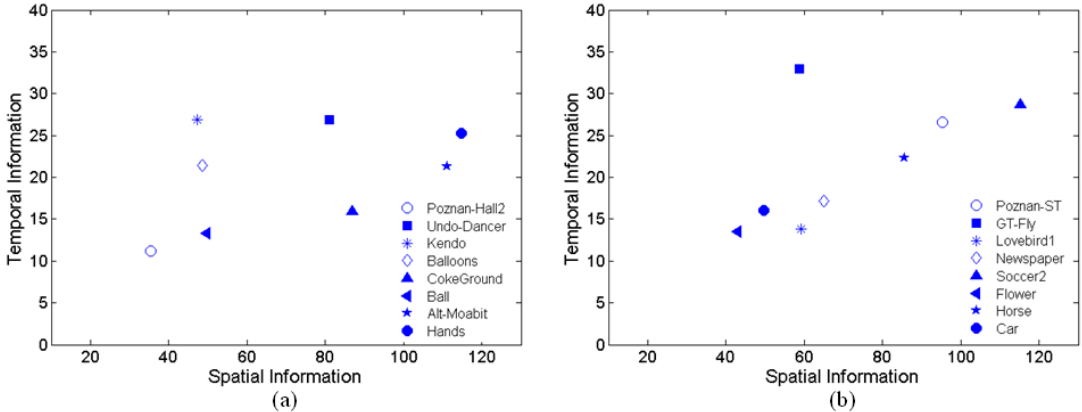

**Fig. 5** Distribution of spatial and temporal information over the training (a) and test (b) datasets

and left views. Then DERS generated a new depth map for the distorted stereo pair [26].

*3)* Shifted (increased) intensity: the brightness intensity of right and left video streams is increased by 20 (out of 255). Once more, a new depth map is generated for the distorted stereo pair using DERS [26].

*4)* High compression of views (simulcast coding): the right and left views are simulcast coded using an HEVC-based encoder (reference software HM ver. 9.2) [46]-[48]. The low delay configuration setting with the GOP (group of pictures) size of 4 was used. The quantization parameter (QP) was set to 35 and 40 to investigate the performance of our proposed metric at two different compression-distortion levels with visible artifacts. A new depth map for the distorted stereo pair is generated using DERS [26].

*5)* High compression of depth map: the depth map of one view (left view in our simulations) is compressed using an HEVC-based encoder (HM reference software ver. 9.2 [46]-[48]) with QPs of 25 and 45, and low delay profile with GOP size of 4. Then, the right view is synthesized using the decoded depth map and the original left view. View synthesis is performed using the view synthesis reference software (VSRS 3.5) [49]. Also, a new depth map for the synthesized view is generated using DERS [26]. The quality of the stereo pair including the synthesized right view and the original left view is compared with that of the reference one.

*6)* High compression of 3D content (views and depth maps): the right and left view video sequences and their corresponding depth maps are encoded using an HEVC-based 3D video encoder (3D HTM reference software ver. 9 [48]) with random access high efficiency configuration. GOP size was set to 8 and QP values were set to 25, 30, 35, and 40 according to the MPEG common test conditions for 3D-HEVC [39].

*7)* View synthesis: using one of the views of each stereo pair and its corresponding depth map, the other view is synthesized via VSRS and a new stereo pair is generated [49]. As a result, two sets of distorted stereo pairs are formed where the synthesized view once was the left view and once the right one. Then, new depth maps are generated for these distorted stereo pairs using DERS [26].

In our study, capturing artifacts (such as window violation, vertical parallax, depth plane curvature, keystone distortion, or shear distortion) are not considered, as our proposed quality metric is a full-reference metric and requires a reference for comparison. In other words, it is assumed that the 3D video content and the corresponding depth map sequences are already properly captured and the goal is to evaluate the perceived quality when other processing/distortions are applied. Distortions introduced during 3D content acquisition using rangefinder sensors are not considered in our experiments. In addition, we did not take into account the effect of crosstalk, as the amount of crosstalk depends on the 3D display technology used. We assume that the display imposes the same amount of crosstalk to both reference and distorted 3D content. To design a quality metric that takes into account the effect of crosstalk, the information about the amount of crosstalk at different intensity level for different 3D display systems is required.

After applying the seven above-mentioned distortions (at different levels) to the sixteen stereo videos, we obtain 208 distorted stereo videos.

*3.2 Subjective test setup*

The viewing conditions for subjective tests were set according to the ITU-R Recommendation BT.500-13 [50]. The evaluation was performed using a 46" Full HD Hyundai 3D TV (Model: S465D) with passive glasses. The peak luminance of the screen was set at 120 cd/m2 and the color temperature was set at 6500K according to MPEG recommendations for the subjective evaluation of the proposals submitted in response to the 3D Video Coding Call for Proposals [51]. The wall behind the monitor was illuminated with a uniform light source (not directly hitting the viewers) with the light level less than 5 % of the monitor peak luminance.

A total of 88 subjects participated in the subjective test sessions, ranging from 21 to 32 years old. All subjects had none to marginal 3D image and video viewing experience. They were all screened for color blindness (using Ishihara chart), visual acuity (using Snellen charts), and stereovision acuity (via Randot test – graded circle test 100 seconds of arc). Subjective evaluations were performed on both training and validation data sets (see Table 1 & Table 2).

Test session started after a short training session, where subjects became familiar with video distortions, the ranking scheme, and test procedure. Test sessions were set up using the single stimulus (SS) method where videos with different qualities were shown to the subjects in random order (and in a different random sequence for each observer). Each test video was 10 seconds long and a four-second gray interval was provided between test videos to allow the viewers to rate the perceptual quality of the content and relax their eyes before watching the next video. There were 11 discrete quality levels (0-10) for ranking the videos, where score 10 indicated the highest quality and 0 indicated the lowest quality. Here, the perceptual quality reflects whether the displayed scene looks pleasant in general. In particular, subjects were asked to rate a combination of "naturalness", "depth impression" and "comfort" as suggested by [52]. After collecting the experimental results, we removed the outliers from the experiments (there were seven outliers) and then the mean opinion scores (MOS) from the remaining viewers were calculated. Outlier detection was performed in accordance to ITU-R BT.500-13, Annex 2 [50].

*3.3 Assigning constant exponents and pooling parameters*

To find the constant exponents in equation (7), the two quality components were calculated for the videos in the training set. Note that in our experiment, a block size of 64×64 was chosen for measuring the variance of disparity. In this case, the 3D display has resolution of 1080×1920 (HD), its height is 773 mm, the appropriate viewing distance is 3000 mm, and the value of $2\alpha$ in equation (4) is equal to 0.88° which is consistent with the fovea visual focus.

Then, the exponent values that result in the highest correlation between the HV3D metric and the MOS values corresponding to the training distorted video sets are selected (see equation (8)). The selected constant exponents are $\beta_1=0.4$, $\beta_2=0.1$, and $\beta_3=0.29$.

In order to find the Minkowski exponent parameter ($p$) and the exponential time constant ($\tau$) for temporal pooling (see equation (9)), the Pearson Correlation Coefficient (which measures the accuracy of a mapping) is calculated for a wide range of these two parameters over the training video set. In other words, these parameters are exhaustively swept over a wide range to enable us find the highest stable maximum point in the accuracy function. Extensive numerical evaluations show that the effect of slight changes in the selected pooling parameters is negligible in the overall metric performance. The same performance evaluations have shown that $p = 9$ and $\tau = 100$ result in the highest correlation between the subjective tests and our quality metric results.

## 4 Results and discussion

In this section we evaluate the performance of each component of the HV3D quality metric (cyclopean view and depth map quality terms) as well as its overall performance over the validation data set. The performance of HV3D is also compared with that of state-of-the-art 2D and 3D quality metrics. The performance of HV3D quality metric is discussed in the following subsections.

*4.1 Contribution of quality components of HV3D metric*

In order to investigate the contribution of the cyclopean view and depth map quality components of HV3D quality metric (see equation 7) in predicting the overall 3D QoE, the correlation between each quality component and the MOS values over the validating dataset is studied by calculating the Spearman rank order correlation coefficient (SCC), the Pearson correlation coefficient (PCC), and the Root-Mean-Square Error (RMSE). While PCC and RMSE measure the accuracy of the 3D QoE prediction by quality components, SCC measures the statistical dependency between the subjective and objective results. In other words Spearman ratio assesses how well the relationship between two variables can be described using a monotonic function, i.e., it measures the monotonicity of the mapping from each quality metric to MOS. In addition, to measure the consistency of mapping from each quality metric component to MOS, outlier ratios (*OR*) are calculated [9]. Table 3 shows SCC, PCC, RMSE, and OR for cyclopean view and depth map quality components of HV3D. As it is observed, the quality components of the proposed metric demonstrate high correlation with Mean Opinion Score (MOS). In the following subsection the overall performance of HV3D as a hybrid combination of the two quality components is analyzed.

*4.2 Overall performance of the HV3D metric*

To prove the efficiency of the HV3D quality metric, its performance is compared with that of the state-of-the-art 3D quality metrics using the validation dataset. We compare our proposed quality metric against Ddl$_1$ [16], OQ [17], CIQ [18], PHVS-3D [19], and PHSD [28]. Note that the parameters of PHVS-3D [19] and PHSD [28] are originally customized for 3D mobile applications. For a fair comparison these parameters are updated for 3D HD screen. In addition, we follow what is considered common practice in evaluating 3D quality metrics and compare the performance of the HV3D against several 2D quality metrics including PSNR, SSIM [13]*,* VIF [25], VQM [53], and MOVIE [54]. To this end, the quality of the frames of both views is measured separately using these 2D quality metrics and then the average quality over the frames from both views is calculated. In order to perform a comprehensive evaluation on the performance of the proposed method, in addition to full-reference 3D and 2D quality metrics, we also include a no-reference quality metric proposed by Ryu and Sohn [24] in our experiments. Table 4 provides an overview on the 3D quality metrics used in our experiments.

**Table 3** Statistical performance of the cyclopean view and depth map quality components of HV3D

| Quality Metric | Spearman Ratio | Pearson Ratio | RMSE | Outlier Ratio |
|---|---|---|---|---|
| $Q_{Cyclopean}$ | 0.8177 | 0.8660 | 7.133 | 0 |
| $Q_{Depth}$ | 0.7993 | 0.8524 | 7.398 | 0 |

**Table 4** Overview of different full-reference stereoscopic quality metrics

| Quality Metric | Direct Use of the Views | Use of Cyclopean View | Use of Depth (Disparity) Map | Combine 2D and 3D Quality | Temporal Pooling | Complexity | Use of Color | Stereoscopic Dataset | Size of the Dataset | Type of Distortions Assessed |
|---|---|---|---|---|---|---|---|---|---|---|
| Ddl1 [16] | Yes | No | Yes | Yes | No | Low | No | Images | Medium | JPEG, JPEG2000, blur |
| OQ [17] | Yes | No | Yes | Yes | No | Low | No | Images | Medium | 2D distortions |
| CIQ [18] | No | Yes | No | No | No | Low | No | Images | Medium | 2D distortions |
| PHVS-3D [19] | No | Yes | No | No | No | High | No | Videos | Small | 3D compression |
| PHSD [28] | No | Yes | Yes | Yes | No | High | No | Videos | Medium | 3D compression |
| MJ3D [21] | No | Yes | No | No | No | Medium | No | Images | Medium | JPEG, JPEG2000, blur, White Noise, Fast Fading |
| Q_Ryu [24] | Yes | No | No | No | No | Medium | Yes | Images & Videos | Large | JPEG, JPEG2000, blur, White Noise, Fast Fading |
| Q_Shao [20] | Yes | No | No | Yes | No | High | No | Images | Medium | JPEG, JPEG2000, blur, White Noise, H.264 |
| Proposed | No | Yes | Yes | Yes | Yes | Medium | No | Videos | Large | 2D & 3D distortions |

Fig. 6 shows the relationship between the MOS and the resulting values from each quality metric for the entire validation set and all 7 different distortions (as described in Section 3). A logistic fitting curve is used for each case to clearly illustrate the correlation between subjective results (MOS) and the results derived by each metric. The logistic fitting curve is formulated as follows [55]:

$$y = \frac{a}{1 + e^{-b(x-c)}} \quad (10)$$

where *x* denotes the horizontal axis (quality metric) and *y* represents the vertical axis (MOS) in each diagram and *a*, *b,* and *c* are the fitting parameters. Fig. 6 shows that our HV3D objective metric demonstrates strong correlation with the MOS results.

In order to evaluate the statistical relationship between each of the quality metrics and the subjective results, PCC, SCC, RMSE, and OR are calculated for different quality metrics over the entire validation dataset and is illustrated in Table 5. As it is observed, the performance of our HV3D in quantifying the quality of the entire validation dataset in the presence of the 7 representative types of distortions is superior to other objective metrics in terms of accuracy, monotonicity, and consistency. In particular, Pearson correlation coefficient between our metric and MOS is 90.8 %, Spearman correlation ratio is 91.3 %, and RMSE is 6.43. As it is observed the hybrid combination of the cyclopean view and depth map quality components has improved the correlation between the quality indices and MOS values (see Table 3 and Table 5).

**Table 5** Statistical performance of different quality metrics over the whole validation dataset

| Quality Metric | Spearman Ratio | Pearson Ratio | RMSE | Outlier Ratio |
|---|---|---|---|---|
| PSNR | 0.6350 | 0.6454 | 10.388 | 0.0167 |
| SSIM [13] | 0.6213 | 0.6844 | 9.852 | 0.0083 |
| VQM [53] | 0.5981 | 0.6660 | 10.095 | 0.0083 |
| VIF [25] | 0.7204 | 0.7257 | 9.166 | 0 |
| MOVIE [54] | 0.7967 | 0.7527 | 8.623 | 0 |
| Ddl1 [16] | 0.7321 | 0.7370 | 8.732 | 0 |
| OQ [17] | 0.7900 | 0.7580 | 8.610 | 0 |
| CIQ [18] | 0.7080 | 0.7200 | 9.446 | 0.0083 |
| PHVS-3D [19] | 0.8233 | 0.7837 | 8.420 | 0 |
| PHSD [28] | 0.7841 | 0.7911 | 8.321 | 0 |
| MJ3D [21] | 0.8947 | 0.8640 | 7.229 | 0 |
| Q_Ryu [24] | 0.8410 | 0.8475 | 7.787 | 0 |
| Q_Shao [20] | 0.7988 | 0.8348 | 7.902 | 0 |
| **HV3D** | **0.9130** | **0.9082** | **6.433** | **0** |

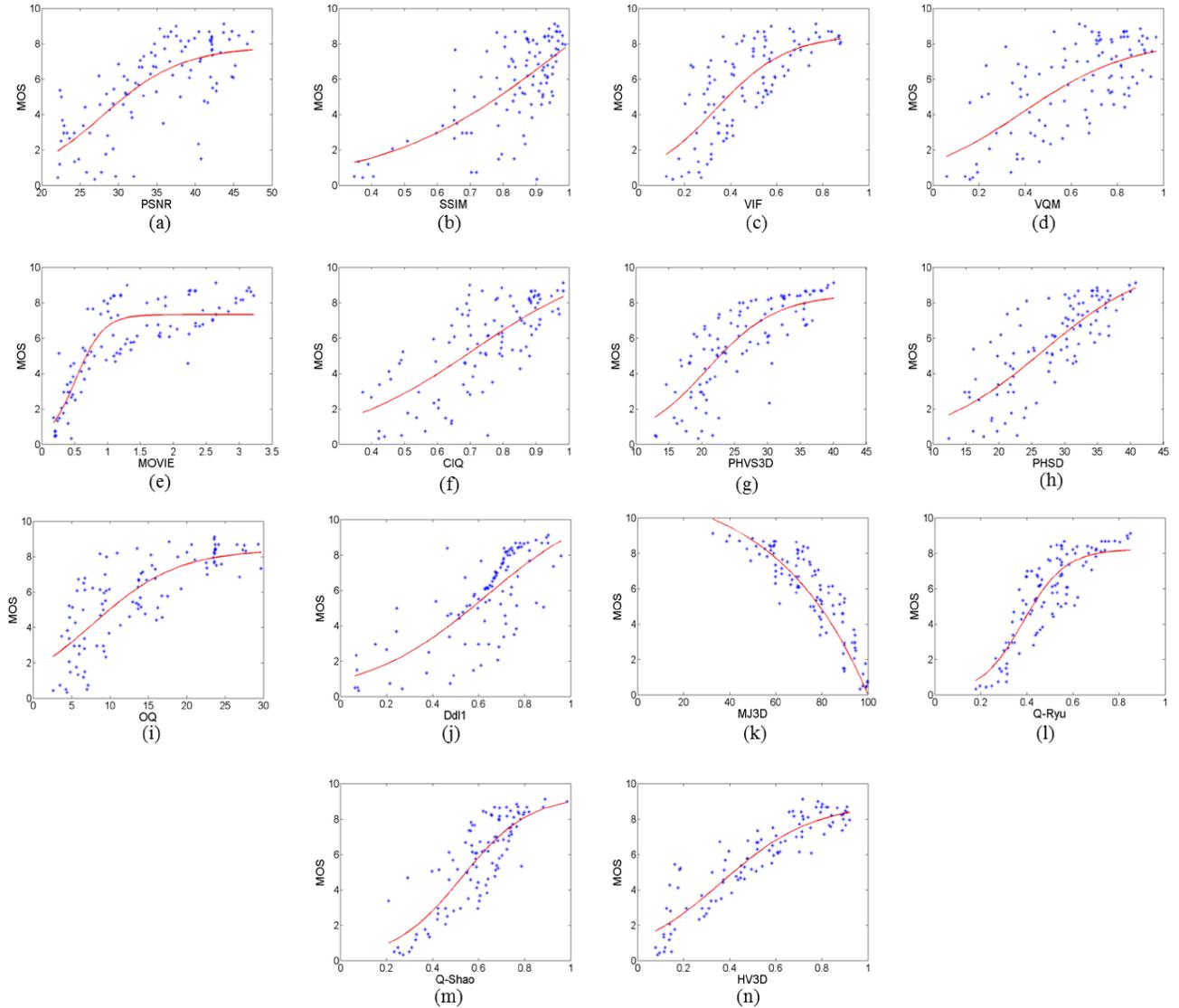

**Fig. 6** Comparing the subjective results with objective results using PSNR (a), SSIM [13] (b), VIF [25] (c), VQM [53] (d), MOVIE [54] (e), CIQ [18] (f), PHVS-3D [19] (g), PHSD [28] (h), OQ [17] (i), Ddl1 [16] (j), MJ3D [21] (k), Q_Ryu [24] (l), Q_Shao [20] (m), and HV3D (n) objective quality metrics

**Table 6** Statistical performance of different quality metrics for each specific type of distortion

| Quality Metric Distortion | Additive Gaussian Noise | | View Compression (Simulcast) | | Blurring | | Brightness Shift | | 3D Video Compression | | View Synthesis | | Depth Map Compression | |
|---|---|---|---|---|---|---|---|---|---|---|---|---|---|---|
| | PCC | SCC | PCC | SCC | PCC | SCC | PCC | SCC | PCC | SCC | PCC | SCC | PCC | SCC |
| PSNR | 0.6832 | 0.6551 | 0.7098 | 0.7211 | 0.5919 | 0.6190 | 0.4931 | 0.4097 | 0.7400 | 0.7318 | 0.6393 | 0.6238 | 0.6503 | 0.6343 |
| SSIM [13] | 0.7716 | 0.7192 | 0.7530 | 0.7480 | 0.7159 | 0.7143 | 0.7653 | 0.7804 | 0.6998 | 0.6530 | 0.6539 | 0.6591 | 0.6700 | 0.6916 |
| VQM [53] | 0.7367 | 0.7120 | 0.6622 | 0.7104 | 0.7496 | 0.7048 | 0.7286 | 0.7687 | 0.6842 | 0.6897 | 0.6934 | 0.6430 | 0.5739 | 0.6178 |
| VIF [25] | 0.7967 | 0.7982 | 0.7694 | 0.7866 | 0.8232 | 0.7619 | **0.8523** | **0.8711** | 0.7307 | 0.7312 | 0.6059 | 0.6548 | 0.7535 | 0.7727 |
| MOVIE [54] | 0.8014 | **0.8189** | 0.7497 | 0.7920 | 0.7564 | 0.7524 | 0.7098 | 0.6916 | 0.7195 | 0.7511 | 0.6877 | 0.7171 | 0.6495 | 0.6462 |
| Ddl1 [16] | 0.6228 | 0.6796 | 0.7620 | 0.8003 | 0.7433 | 0.7190 | 0.7761 | 0.7152 | 0.7200 | 0.7311 | 0.7255 | 0.7138 | 0.8132 | 0.8197 |
| OQ [17] | 0.7133 | 0.7347 | 0.6859 | 0.7425 | 0.7042 | 0.7714 | 0.6519 | 0.7074 | 0.7411 | 0.7805 | 0.6816 | 0.7196 | 0.7071 | 0.7696 |
| CIQ [18] | 0.7769 | 0.7868 | 0.7741 | 0.7552 | 0.8325 | 0.8238 | 0.7240 | 0.7289 | 0.7556 | 0.7410 | 0.7123 | 0.7315 | 0.7701 | 0.8447 |
| PHVS-3D [19] | 0.6918 | 0.6791 | 0.7960 | 0.8039 | 0.7244 | 0.7286 | 0.6079 | 0.6048 | 0.8194 | 0.8430 | 0.7366 | 0.7197 | 0.7840 | 0.7711 |
| PHSD [28] | 0.6484 | 0.6755 | 0.8522 | **0.8789** | 0.7523 | 0.7048 | 0.6249 | 0.6166 | 0.8330 | 0.8572 | 0.7532 | 0.7746 | 0.8285 | 0.8099 |
| MJ3D [21] | **0.8277** | 0.8098 | 0.8452 | 0.8467 | 0.8123 | 0.8426 | 0.8001 | 0.8319 | 0.8009 | 0.8234 | 0.7219 | 0.7341 | 0.7018 | 0.7422 |
| Q_Ryu [24] | 0.7611 | 0.7776 | **0.8671** | 0.8523 | **0.8612** | **0.8549** | 0.6912 | 0.7461 | 0.8081 | 0.8233 | 0.6729 | 0.6644 | 0.7511 | 0.7636 |
| Q_Shao [20] | 0.7988 | 0.7812 | 0.8233 | 0.8122 | 0.8278 | 0.8167 | 0.7012 | 0.6911 | 0.7923 | 0.7786 | 0.7088 | 0.7121 | 0.7245 | 0.7098 |
| HV3D | 0.7994 | 0.7823 | 0.8312 | 0.8466 | 0.8108 | 0.8001 | 0.8412 | 0.8539 | **0.8965** | **0.9010** | **0.8881** | **0.8443** | **0.8603** | **0.8554** |

*4.3 Performance of HV3D in the presence of different types of distortions*

To evaluate the performance of the proposed metric in predicting the quality of the content in the presence of different types of distortions, the statistical relationship between HV3D indices and MOS values is analyzed separately per each type of distortion.

Table 6 shows the PCC and SCC values for various quality metrics and different types of distortions over the validation dataset. As it is observed the HV3D quality metric either outperforms other quality metrics or its performance is quite comparable in predicting the quality of distorted 3D content. The results in Table 6 show superior performance of our proposed quality metric specifically for 3D video coding and view synthesizing applications. Note that in the case of simulcast video compression the performance of our quality metric is slightly lower than the 3D video compression case. This is due to the low quality of depth maps generated from the compressed stereo videos in the simulcast coding scenario (in 3D video coding the coded version of reference depth maps are available).

*4.4 Effect of the temporal pooling*

The performance evaluations demonstrated in Table 5 presents statistical comparison between HV3D and various 2D and 3D quality metrics. Except VQM [53], MOVIE [54], and HV3D, the rest of the metrics in Table 5 have been originally designed for assessing the quality of images and do not take into account the temporal aspect of video content. To address this, the exponentially weighted Minkowski pooling mechanism [37] is used in conjunction with image quality metrics to convert a set of frame quality scores to a single meaningful score for a video instead of averaging the scores over the frames. Note that the parameters of the exponentially weighted Minkowski pooling are optimized for each quality metric separately (using the training set), to achieve the highest PCC with the MOS scores.

The PCC and SCC values between the MOS and different quality metrics before and after applying the temporal pooling are reported in Table 7. As it is observed temporal pooling in general tends to improve the performance of the metrics. This improvement is more substantial in the case of the quality metrics such as PSNR and SSIM that have lower assessment performance at the presence of distortions with more temporally variant visual artifacts such as 3D compression, view synthesis, and depth map compression.

It is observed from Table 7 that, in general, temporal pooling does not have a significant impact on the performance of different quality metrics (maximum effect occures for PSNR, which is an increase in SCC of 2.53 % and an icrease in PCC of 2.07 %). The reason is that the distortions added to the original videos in Section 3.1 (which are the common artifacts that occur during compression, transmission, and display of 3D content) are not highly variant over time. For this reason, the video database contains videos in which distortion densities do not change drastically from frame to frame. As a result, the visual quality of different parts of the videos does not vary significantly over time. In order to investigate the effect of temporal pooling when distortions are highly variant in time, we perform additional tests using a new set of videos containg time-variant distortions. Note that simulcast compression, depth map compression, 3D video compression, and view synthesis distortions are not applied on each frame independently and are almost visually consistent over the entire video sequence. For that reason, we do not consider these distortions in this experiment. The types of artifacts that are considered are the ones which show spiky appearance in some frames and do not appear in the rest of frames. The following distortions are considered in our additional experiment:

*1)* White Gaussian noise: white Gaussian noise with zero mean and variance value varying temporally from 0.01 to 0.08 is applied to both right and left views; DERS is used to generate a new depth map for the distorted stereo pair [26]. The variance value for each frame is selected randomly between 0.01 and 0.08.

*2)* Gaussian low pass filter: a Gaussian low pass filter with the size $L_G$ and standard deviation of $S_G$ is applied to both right and left views. Then, DERS is used to generate a new depth map for the distorted

**Table 7** Statistical performance of different quality metrics with/without temporal pooling

| Quality Metric | Averaging the frame quality scores | | With Temporal Pooling | |
|---|---|---|---|---|
| | SCC | PCC | SCC | PCC |
| PSNR | 0.6350 | 0.6454 | 0.6603 | 0.6661 |
| SSIM [13] | 0.6213 | 0.6844 | 0.6405 | 0.7019 |
| VQM [53] | NA | NA | 0.5981 | 0.6660 |
| VIF [25] | 0.7204 | 0.7257 | 0.7239 | 0.7422 |
| MOVIE [54] | NA | NA | 0.7967 | 0.7527 |
| Ddl1 [16] | 0.7321 | 0.7370 | 0.7325 | 0.7501 |
| OQ [17] | 0.7900 | 0.7580 | 0.7911 | 0.7790 |
| CIQ [18] | 0.7080 | 0.7200 | 0.7140 | 0.7216 |
| PHVS-3D [19] | 0.8233 | 0.7837 | 0.8247 | 0.7862 |
| PHSD [28] | 0.7841 | 0.7911 | 0.7946 | 0.7996 |
| MJ3D [21] | 0.8947 | 0.8640 | 0.8988 | 0.8691 |
| Q_Ryu [24] | 0.8410 | 0.8475 | 0.8521 | 0.8538 |
| Q_Shao [20] | 0.7988 | 0.8348 | 0.7999 | 0.8401 |
| HV3D | **0.9014** | **0.8959** | **0.9130** | **0.9082** |

**Table 8** Statistical performance of different quality metrics with/without temporal pooling when the distortions densities are highly variant in time

| Quality Metric | Averaging the frame quality scores | | With Temporal Pooling | |
|---|---|---|---|---|
| | SCC | PCC | SCC | PCC |
| PSNR | 0.4075 | 0.3801 | 0.5098 | 0.4745 |
| SSIM [13] | 0.6047 | 0.6571 | 0.6816 | 0.7182 |
| VQM [53] | NA | NA | 0.4989 | 0.5355 |
| VIF [25] | 0.6822 | 0.7566 | 0.7678 | 0.7931 |
| MOVIE [54] | NA | NA | 0.7599 | 0.7783 |
| Ddl1 [16] | 0.6946 | 0.7169 | 0.7472 | 0.7618 |
| OQ [17] | 0.7209 | 0.7481 | 0.7889 | 0.7999 |
| CIQ [18] | 0.6917 | 0.6378 | 0.7223 | 0.6949 |
| PHVS-3D [19] | 0.7511 | 0.7693 | 0.7969 | 0.8032 |
| PHSD [28] | 0.7703 | 0.7868 | 0.8057 | 0.8218 |
| MJ3D [21] | **0.8081** | 0.7995 | 0.8410 | 0.8357 |
| Q_Ryu [24] | 0.7590 | 0.7874 | 0.8089 | 0.8219 |
| Q_Shao [20] | 0.7056 | 0.7659 | 0.7777 | 0.8119 |
| HV3D | 0.8055 | **0.8099** | **0.8559** | **0.8403** |

stereo pair [26]. $L_G$ and $S_G$ vary randomly in the interval of [3-15] for each frame.

*3)* Shifted (increased) intensity: the brightness intensity of the right and left frames is increased by randomly selecting pixel values from [10-50] (out of 255). A new depth map is generated for the distorted stereo pair using DERS [26].

*4)* JPEG compression of the views: both views are compressed using JPEG compression with the quality factor randomly selected from the range [1-100] for each frame.

Subjective experiments (similar to the ones explained in Section 3.2, for the 4×8 above test videos and 16 subjects) are performed to evaluate the performance of different metrics before and after temporal pooling. Table 8 shows the results of this experiment. It is observed that temporal pooling implies a much stronger effect in the case that distortions densities are highly variant in time. In particular, the maximum effect introduced by temporal pooling occures for PSNR, which is a 10.23 % increase in SCC and 9.44 % in PCC. Moreover, the average improvements in Table 7 due to temporal pooling are 0.79 % for SCC and 1.09 % for PCC, while the improvements in Table 8 are 5.85 % for SCC and 4.68 % for PCC.

*4.5 Sensitivity of our method to constant exponents*

The method we presented in this paper is based on training our quality model using a training video set and then testing it over a validation video set. For these kinds of approaches to be convincing, it is necessary to study the robustness of the algorithm against the changes in the parameters that are used in the design.

In order to evaluate the robustness of the proposed quality metric to the constant exponents, $β_1$, $β_2$, and $β_3$, we first choose a set of optimal parameters e.g. the ones reported in section 3.3. Then, we evaluate the PCC values when the exponents are slightly changed.

Over a wide range of $β_1$, $β_2$, and $β_3$, we observed that the rate of change (derivative) of PCC is not significant. In particular, these parameters are swept over the intervals $0.35 ≤ β_1 ≤ 0.45$, $0.05 ≤ β_2 ≤ 0.15$, and $0.25 ≤ β_3 ≤ 0.35$. Over these intervals, the PCC does not degrade significantly. Minimum PCC value for this set of intervals is 0.8533 (corresponding to SCC value of 0.8572) which occurs at $β_1=0.45$, $β_2=0.15$, and $β_3=0.35$.

*4.6 Complexity of HV3D*

Identifying matching areas over the right and left views is considered to be one of the most computationally complex procedures in the HV3D implementation. As explained in Section 2.1, to model the cyclopean view for each block within one view, the matching block within the other view is detected through an exhaustive search. As illustrated in Fig. 2, $A_R$ is the matching block for $A_L$, which is found by performing an exhaustive search around $A_D$. To reduce the complexity of HV3D, instead of performing a full search, the disparity map is used to find the matching areas within two views. To this end, for each block in the left view, we assume that the horizontal coordinate of the matching block in the right view is equal to the coordinate of the block in the left view plus the disparity of the block in the left view. This approximation as shown in Fig. 2 is as if $A_D$ in the right view is chosen as the matching block for $A_L$ in the left view. Using this approach, which is called Fast-HV3D, the computational complexity is reduced significantly. The experimental results show that for "Fast-HV3D" method, PCC is 0.8865, SCC is 0.8962, RMSE is 6.73, and the outlier ratio is 0, which confirms that while the complexity of the Fast-HV3D quality metric is less than that of the original HV3D quality metric, its performance is almost similar (see Table 5).

Note that due to the use of the disparity information for modeling the binocular fusion of the two views in Fast-HV3D, the cyclopean view quality component will be slightly correlated to the depth map quality component. However, the disparity information is used in a different way in the two quality components. More specifically, the disparity information incorporated in the cyclopean view generation is only used to find the matching blocks and the goal is to fuse the two views to create the intermediate cyclopean image. However, the depth information used in the depth map quality component is directly used to consider: 1) the effect of smooth/fast variations of depth level, and 2) the effect of distortions on the depth map quality. In addition, please note that the depth map quality component does not utilize the image intensity values at all. The cyclopean view image is a 2D intermediate image, constructed by the fusion of the two views. The depth map specifies how much each object (within the cyclopean view) is perceived outside/inside the display screen. It is observed from Table 3 and Table 5 that each quality component can predict the overall MOS to some extent. However, only the combination of the two components demonstrates very high accuracy.

Note that for the Fast-HV3D method, the performance of the metric is slightly less than regular HV3D, which is due to the usage of median disparity for each block and the resulting correlation between the two quality components.

The computational complexity of different algorithms is usually expressed by the complexity degree order, which is mathematically measured. Mathematical measurement of the computational complexity of various quality metrics in terms of complexity degree order is very difficult. Thus to compare the complexity of different quality metrics, we measure the simulation time for each metric. A comparative experiment was performed on a Win7-64bit Workstation, with Intel Core i7 CPU, and 18 GBs of memory. During the experiment, it was ensured that no other program was running on the machine. Each metric was applied to a number of frames, and the total simulation time was measured. Moreover the simulation time of each metric relative to that of PSNR was calculated as follows:

**Table 9** Complexity of different quality metrics

| Quality Metric | Average Simulation Time per One HD Frame (seconds) | Simulation Time relative to PSNR Simulation Time |
|---|---|---|
| PSNR | 0.7 | 1 |
| SSIM [13] | 1.2 | 1.71 |
| VQM [53] | 65.3 | 93.29 |
| VIF [25] | 6.9 | 9.86 |
| MOVIE [54] | 3480.6 | 4972.29 |
| Ddl1 [16] | 1.3 | 1.86 |
| OQ [17] | 1.3 | 1.86 |
| CIQ [18] | 2.3 | 3.29 |
| PHVS-3D [19] | 317.4 | 453.43 |
| PHSD [28] | 323.4 | 462 |
| MJ3D [21] | 46.1 | 65.86 |
| Q_Ryu [24] | 33.02 | 47.17 |
| Q_Shao [20] | 260.33 | 371.9 |
| Fast-HV3D | 31.2 | 44.57 |
| HV3D | 39.6 | 56.57 |

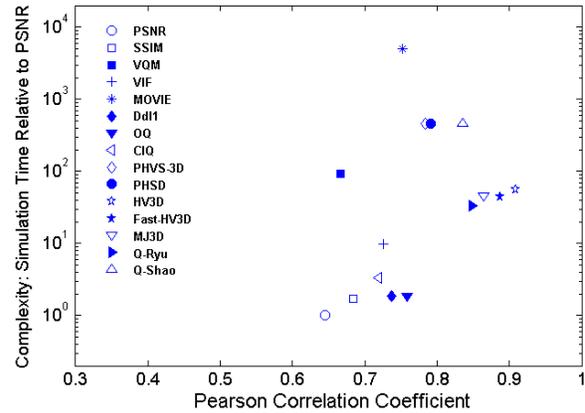

**Fig. 7** Relative complexity of different metrics versus their PCC values

$$\text{Simulation time relative to } PSNR = \frac{\text{Simulation time for each metric}}{PSNR \text{ simulation time}} \tag{11}$$

The average simulation times per one HD frame for different quality metrics, as well as the relative simulation time of each quality metric with respect to that of PSNR are reported in Table 9. As it is observed, the complexity of Fast-HV3D is 21.21% less than that of HV3D in terms of simulation time. Although HV3D is 56.5 times more complex than PSNR in terms of simulation time, its complexity is still much less than that of other quality metrics such as PHSD, PHSD-3D or MOVIE, which are 453.43 to 4972.29 times more complex than PSNR. The relative simulation time versus the PCC value for different quality metrics is illustrated in Fig. 7. It is observed that the proposed HV3D quality metric and its fast implementation (Fast-HV3D) perform well, while their computational complexity is moderate.

In summary, the performance evaluations and Spearman and Pearson correlation ratio analysis showed that our 3D quality metric quantifies the degradation of quality caused by several representative types of distortions very competitively compared to the state-of-the-art quality metrics. In our future work, we will improve the temporal pooling approach used in HV3D by including other factors such as motion, depth of the scene, and frame rate.

## 5 Conclusions

In this paper we proposed a new full-reference quality metric called HV3D for 3D video applications. Our approach models the human stereoscopic vision by fusing the information of the left and right views through 3D-DCT transform and takes to account the sensitivity of the human visual system to contrast as well as the depth information of the scene. In addition, a temporal pooling mechanism is utilized to account for the temporal variations in the video quality.

To adjust the parameters of the HV3D quality metric and evaluate its performance, we prepared a database of 16 reference and 208 distorted videos with representative types of distortions. Performance evaluations revealed that the proposed quality metric achieves an average of 90.8 % correlation between HV3D and MOS, outperforming the state-of-the-art 3D quality metrics. The proposed metric can be tailored to different applications, as it takes into account the display size (the distance of the viewer from the display) and the video resolution.